# Dissipative quadratic soliton mode-locking of nonlinear frequency conversion

Jonathan Musgrave[1] Mingming Nie[1,2] and Shu-Wei Huang[1,*]

[1]Department of Electrical, Computer and Energy Engineering, University of Colorado Boulder, Boulder, Colorado 80309, USA
[2]Key Laboratory of Optical Fiber Sensing and Communications (Education Ministry of China), University of Electronic Science and Technology of China, Chengdu 611731, China
[*]Corresponding author: shuwei.huang@colorado.edu

**Abstract:** Nonlinear frequency conversion underpins numerous classical and quantum photonics applications but conventionally relies on synchronized femtosecond mode-locked lasers and dispersion-engineered enhancement cavities—an approach that imposes substantial system complexity. Here, we report a fundamentally different paradigm: dissipative quadratic soliton (DQS) mode-locking in a continuous-wave (CW)-pumped, doubly resonant second-harmonic generation cavity. By leveraging a cascaded quadratic nonlinear process, we realize an effective Kerr nonlinearity (EKN) that exceeds the intrinsic material Kerr response by over three orders of magnitude and is tunable in both magnitude and sign via pump detuning. This engineered nonlinearity enables femtosecond DQS formation in a free-space lithium niobate cavity with normal dispersion, without dispersion engineering or synchronization electronics. Numerical simulations predict distinct dynamical regimes depending on phase detuning, and experiments confirm the spontaneous emergence of bichromatic femtosecond solitons spanning visible and near-infrared wavelengths. The observed DQSs exhibit spectral 3 dB bandwidths and transform-limited pulse durations of 1.15 THz and 274 fs for the pump and 1.13 THz and 279 fs for the second harmonic. Our results establish a versatile platform for efficient and broadband nonlinear frequency conversion and frequency comb generation based on quadratic nonlinearities, with significant implications for scalable ultrafast and nonlinear photonics applications.

## Introduction

Nonlinear frequency conversion lies at the heart of modern photonics, enabling the generation and manipulation of light at wavelengths inaccessible to conventional sources [1]. By harnessing quadratic nonlinear optical interactions such as second-harmonic generation (SHG), sum- and difference-frequency generation (SFG and DFG), and optical parametric oscillation, researchers have developed highly tunable light sources critical for applications such as precision metrology, field manipulation, and ultrafast spectroscopy [2–11]. These processes in the quantum regime also serve a foundational role in quantum information science, enabling spectral bridging between disparate systems and facilitating entanglement distribution across quantum networks [12–21].

Traditionally, cavity-enhanced nonlinear frequency conversion has been employed to boost the efficiency of inherently weak nonlinear processes by increasing the effective interaction length and intracavity power. This is typically achieved using a dispersion-free resonant cavity synchronously pumped by a femtosecond mode-locked laser (MLL) [2–11]. However, the reliance on femtosecond MLL and the need for precise synchronization electronics introduce substantial system complexity, increasing the physical footprint, and raise the overall cost, posing challenges for widespread adoption.

Dissipative Kerr solitons (DKSs) have recently emerged as a promising alternative to femtosecond MLLs, offering a simplified approach to ultrafast light generation [22–27]. These self-localized waveforms can form spontaneously in continuous-wave (CW) laser-driven passive resonators when a delicate double balance is achieved among Kerr nonlinearity, dispersion, cavity loss, and parametric gain. While the DKS approach substantially reduces the complexity of ultrafast lasers, its application to cavity-enhanced nonlinear frequency conversion remains limited. First, due to the intrinsically weak Kerr nonlinearity, DKS formation has not been observed in free-space cavities, which lack the strong mode confinement provided by fibers or microresonators. Second, Kerr nonlinearity does not support large-offset frequency conversion, unlike quadratic (second-order) nonlinear processes. As a result, systems that rely on efficient nonlinear frequency conversion still require enhancement cavities based on quadratic nonlinearity and the use of precise synchronization electronics.

In this work, we propose and demonstrate mode-locking of nonlinear frequency conversion by leveraging the physics of dissipative quadratic soliton (DQS) [28–31]. Identifying the similarities between DKS and DQS, we have predicted that femtosecond DQS can also form spontaneously in a CW-pumped quadratic nonlinear enhancement cavity. Here, we experimentally validate the prediction and achieve DQS mode-locking of a CW-pumped doubly resonant cavity enhanced SHG in free space, resulting in bichromatic frequency combs spanning visible and near-infrared. Our results establish a new route toward efficient and broadband nonlinear frequency conversion without the complexity of synchronized femtosecond MLLs, offering a scalable and robust alternative platform for applications in both classical and quantum photonics.

While DQS and parametrically driven DKS [22–24] share certain similarities, it is crucial to highlight their fundamentally different mechanisms for balancing dispersion and nonlinearity, as well as the implications of this distinction. In parametrically driven DKS, dispersion engineering is employed to create anomalous dispersion that balances the intrinsically positive material Kerr nonlinearity (MKN). In contrast, DQS relies on an effective Kerr nonlinearity (EKN) arising from cascaded quadratic processes, which not only dominates over MKN but can also be readily tuned via pump detuning (see Operating Principle) [28]. Notably, EKN reaches its maximum near zero pump detuning and reverses sign when detuning is tuned across this point (Fig. 2a). This tunability enables EKN to compensate for either normal or anomalous dispersion, offering a wide design space for soliton formation. Such nonlinearity engineering provides enhanced flexibility in choosing the soliton spectral range and enables the possibility of universal soliton existence. As a demonstration, our free-space doubly resonant enhancement cavity (Fig. 1) - which contains no dispersion engineering elements and is dominated by the normal material dispersion of periodically poled lithium niobate (PPLN) - successfully generates bichromatic solitons (Fig. 4), made possible by the engineered EKN in the DQS regime.

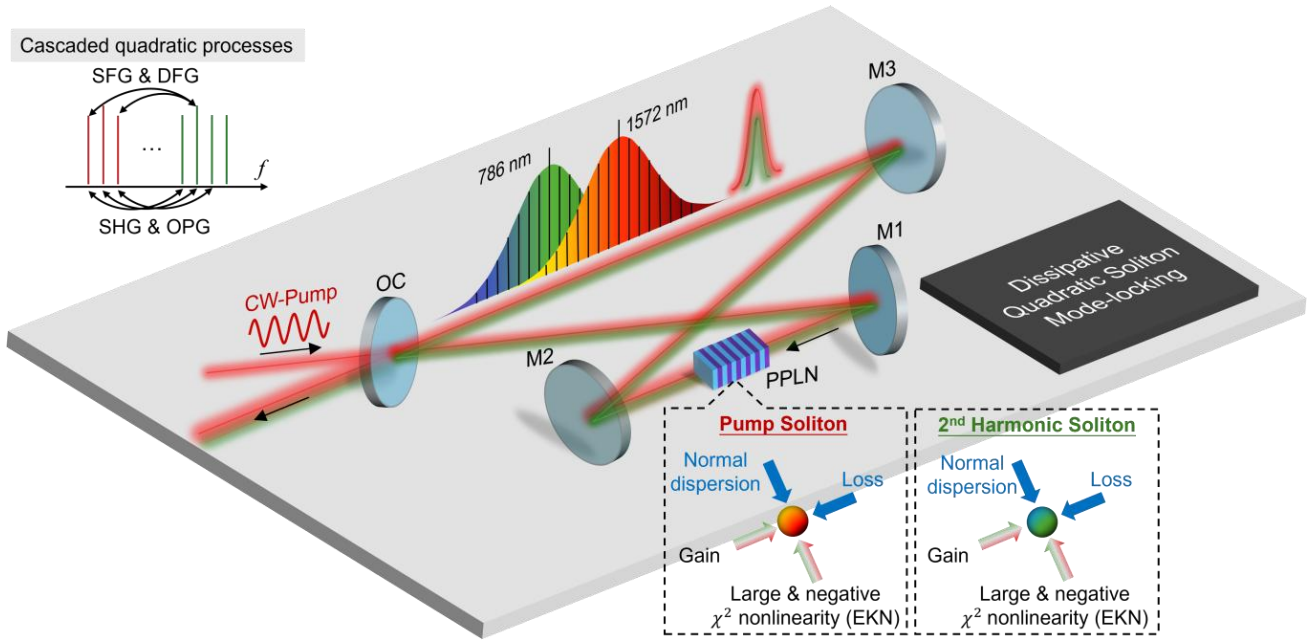

**Fig. 1 DQS mode-locked cavity enhanced second harmonic generation.** The four-mirror free-space cavity incorporating a PPLN crystal is pumped by a continuous wave laser with *p* polarization for type-I phase matching. Top left corner: the cascaded quadratic processes. Bottom right corner: The balance of the cavity effects (blue arrows) and the quadratic nonlinear effects (red and green arrows) to generate the bichromatic solitons at the pump (left inset) and 2$^{nd}$ harmonic (right inset). OC: 1% output coupler at 1572 nm and 786 nm, M1-M3: mirrors with high reflectivity at 1572 nm and 786 nm.

## Results

**Operating principle.** Fig. 1 illustrates the DQS mode-locking principle of the CW-pumped doubly resonant cavity enhanced SHG. As the CW pump - corresponding to the fundamental field - accumulates in the cavity, the second-harmonic (SH) field power increases and eventually saturates. This saturation initiates a series of cascaded quadratic nonlinear interactions, including SHG, SFG, DFG, and optical parametric generation (OPG). Collectively, these processes emulate an effective four-wave mixing mechanism, enabling the spontaneous formation of self-localized DQS, analogous to the dynamics underlying DKS generation.

Section II of the Supplementary Information details the theory of the field evolution in the CW-pumped doubly resonant cavity enhanced SHG [32]. Importantly, when the wave-vector mismatch $\Delta k'$ and group velocity mismatch (GVM) between the pump and SH fields are both zero, the coupled equations describing the DQS dynamics can be simplified to a single mean-field equation for the pump in normalized units [28]:

$$t_R \frac{\partial A}{\partial t} = \left(-\alpha_1 - i\delta_1 - i\frac{k_1'' L}{2}\frac{\partial^2}{\partial \tau^2}\right) A - \alpha_{\text{TPA}} L |A|^2 A + i\gamma_{eff} L |A|^2 A + \sqrt{\theta_1} A_{in}.$$

(1)

Here $t$ is the "slow time" that describes the envelope evolution over successive roundtrips, $t_R$ is the roundtrip time, $\tau$ is the "fast time" that depicts the temporal profiles in the retarded time frame, and $\alpha_1$ is the total linear cavity loss for the pump, $\delta_1$ is the pump phase detuning, $k_1''$ is the pump group velocity dispersion (GVD), $L$ is the nonlinear cavity length, $\theta_1$ is the pump coupling transmission coefficient, $|A_{in}|^2$ is the CW pump power. $\alpha_{TPA} = \alpha_2 L \kappa^2 /(\delta_2^2 + \alpha_2^2)$ is the effective two photon absorption (ETPA) coefficient and $\gamma_{eff} = \delta_2 L \kappa^2/(\delta_2^2 + \alpha_2^2)$ is the effective Kerr nonlinearity (EKN) coefficient. $\alpha_2$ is the total linear cavity loss for the SH and $\delta_2 = 2\delta_1$ is the SH phase detuning. $L$ is the nonlinear length and $\kappa$ is the normalized second-order nonlinear coupling coefficient.

Figure 2a shows the dependence of ETPA and EKN with the pump phase detuning. Notably, as the pump phase detuning increases, the ETPA approaches zero, and equation (1) reduces to the Lugiato–Lefever equation (LLE), which governs conventional DKS dynamics [25]. Due to the large quadratic nonlinearity, the EKN can exceed the intrinsic material Kerr nonlinearity by more than three orders of magnitude in a 10-mm-long type-I phase-matched PPLN crystal. This substantial enhancement significantly lowers the pump power threshold, enabling DQS mode-locking in a free-space cavity. More importantly, the sign of the EKN is governed by the pump phase detuning, allowing it to compensate for either normal or anomalous dispersion and thereby providing a broad design space for DQS formation.

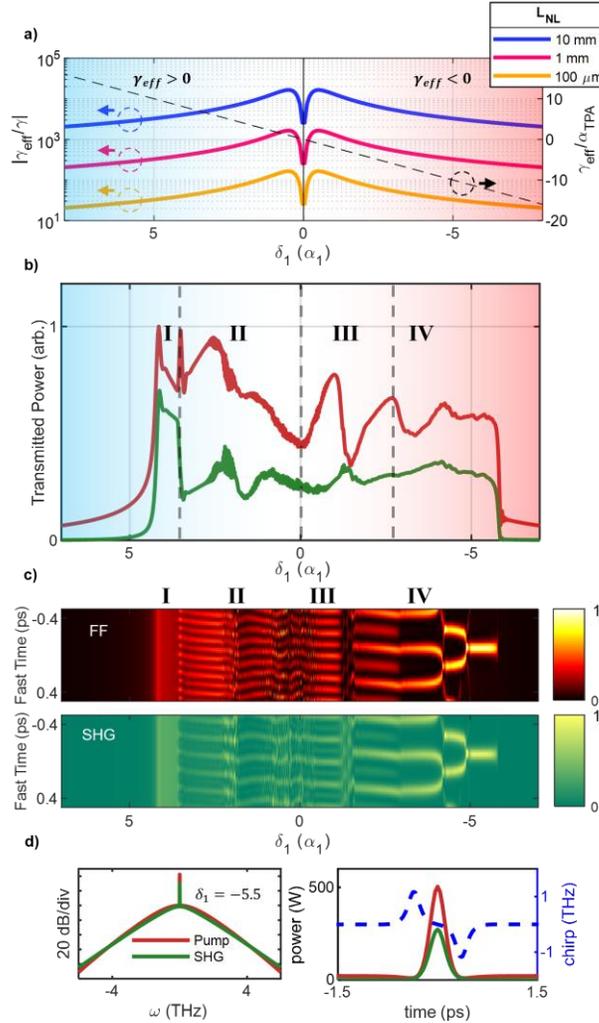

**Fig. 2 Numerical simulation of DQS mode-locking** – a) Ratio of EKN to MKN (left-axis) and EKN to ETPA (right-axis) as a function of pump phase detuning. b) Simulated cavity transmission as the pump frequency is swept over the cavity resonance from blue to red sides. c) The corresponding time-domain evolution of the pump and SH fields reveals several distinct dynamical states, which can be categorized into four types. I) CW state, II) upper-zone temporal patterns, III) lower-zone temporal patterns, and IV) lower-zone DQS states in which the soliton number decreases monotonically as the pump phase detuning becomes more negative. d) Single DQS optical spectra and pulse shapes when $\delta_1 = -5.5\,\alpha_1$ showing a 3-dB bandwidth and pulse duration of 1.38 THz and 263 fs for the pump and 1.08 THz and 276 fs for the second harmonic.

## Numerical simulation of DQS mode-locking

Equation 1 highlights the connection to DKS physics and offers physical insight into the conditions for soliton existence. To comprehensively investigate the DQS mode-locking dynamics in a CW-pumped, doubly resonant cavity-enhanced SHG system, we numerically solved the coupled equations S1–S4, as illustrated in Figure 2. The simulation parameters listed in Table S1 closely reflect the actual experimental conditions.

Figure 2b shows the simulated cavity transmission as the pump frequency is swept over the cavity resonance from blue to red sides and Figure 2c shows the corresponding time-domain evolution of the pump and SH fields reveals several distinct dynamical states, which can be categorized into four types. State I is trivial when the CW pump accumulates in the cavity and the SH field power increases before saturation. Next, the SHG saturation initiates a series of cascaded quadratic nonlinear interactions that result in the transition into state II. Here the pump phase detuning remains positive, and the corresponding upper-zone EKN and ETPA have spectral anomalies (Figure S2) that triggers the spontaneous temporal pattern formation [28].

Of note, the PPLN, and consequently the cavity, has normal dispersion at both pump and SH wavelengths and thus a lower-zone DQS with bichromatic $sech^2$ spectra is expected when the pump phase detuning is swept into the negative regime (red detuning side) [28]. When the pump phase detuning is slightly negative and close to zero, the lower-zone EKN is highly dispersive (Figure S2), and the EKN-to-ETPA ratio is low (Figure 2a). This results in soliton disruption and unstable temporal patterns, characteristic of state III. As the pump frequency is further red-detuned, the EKN becomes less dispersive, and the EKN-to-ETPA ratio increases, leading to the formation of stable, temporally trapped bichromatic DQSs, corresponding to state IV. In this regime, the number of DQSs decreases monotonically with increasing negative detuning. Ultimately, at a pump phase detuning around $\delta_1 = -5.5\,\alpha_1$, a single DQS is generated with a 3-dB bandwidth and pulse duration of 1.38 THz and 263 fs for the pump and 1.08 THz and 276 fs for the second harmonic.

In addition, the perturbative effects of phase mismatch, GVM, and GVD are discussed in Sections III, IV, and V of the Supplementary Information, respectively. Section VI then presents simulation results for a cavity with anomalous dispersion, where DQS mode-locking occurs in the positive pump phase detuning regime instead. These findings underscore the tunability of the EKN, which can be tailored to compensate for either normal or anomalous dispersion, thereby providing a broad design space for soliton formation.

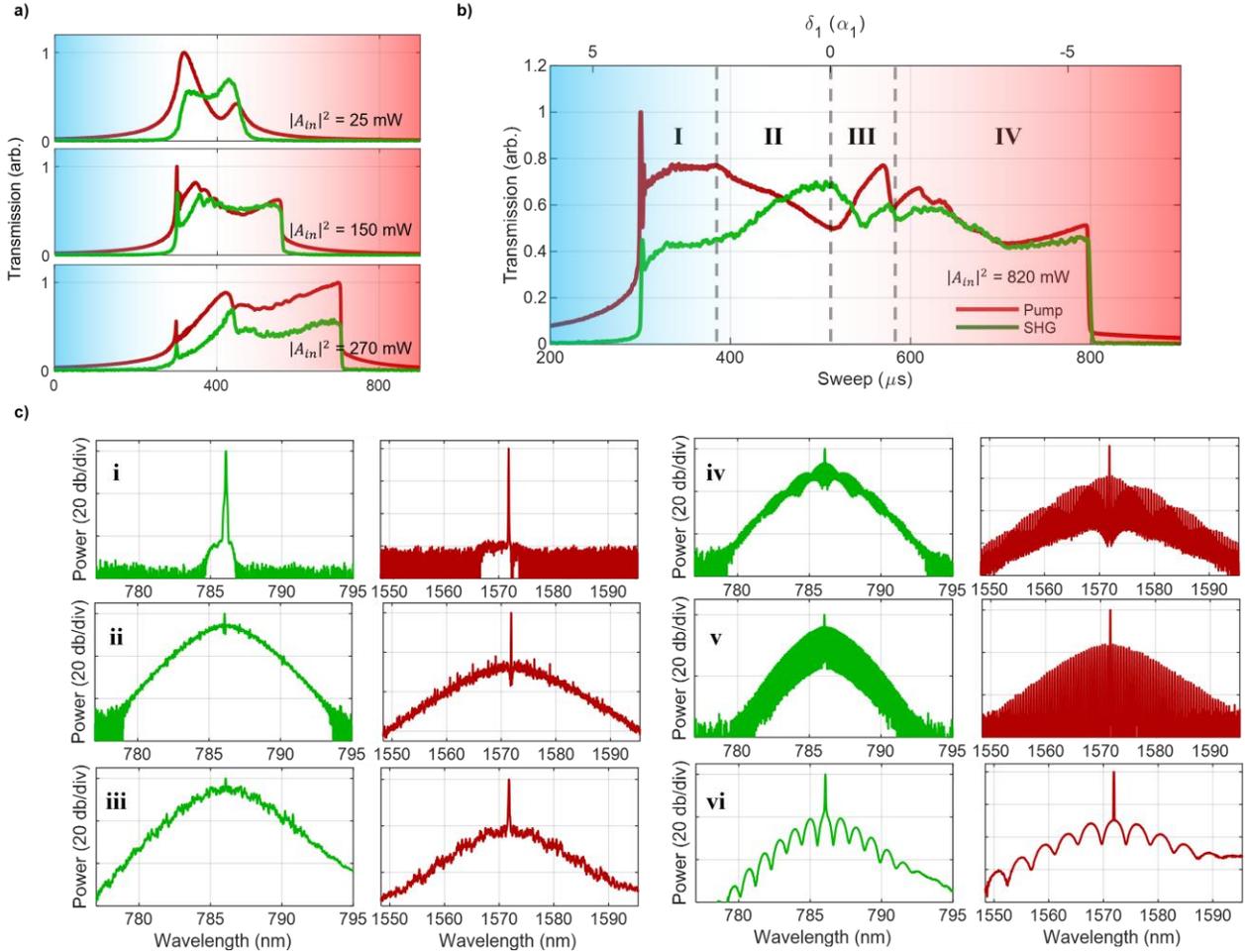

**Fig. 3 Experimental demonstration of DQS mode-locking** – a) Cavity transmission of the pump (red) and SH (green) at different pump power as the pump frequency is swept over the cavity resonance from blue to red sides. b) With an 820-mW pump power, step-like features in the cavity transmission are observed, corresponding to distinct dynamical states that agree with numerical simulations. c) Optical spectra at the pump and SH wavelengths of different dynamical states. i) CW state I, ii) upper-zone temporal pattern state II, iii) lower-zone temporal pattern state III, and iv-vii) lower-zone DQS states with different soliton numbers.

**Experimental demonstration of DQS mode-locking.** We constructed a CW-pumped, doubly resonant cavity-enhanced SHG system with a free spectral range (FSR) of 390 MHz and a loaded quality factor of $2.70 \times 10^8$ (Figure 1) to experimentally demonstrate DQS generation. The pump wavelength was set to 1572 nm to achieve zero GVM in type-I phase-matched PPLN. While type-I phase matching reduces the effective nonlinear coefficient ($d_{eff}$) to 2.7 pm/V, it enables birefringence-based tuning to satisfy both phase matching and doubly resonant conditions via precise temperature control of the PPLN crystal (with 10 mK resolution) [33]. The GVD at the pump and SH wavelengths are 105 fs²/mm and 378 fs²/mm, respectively, which precludes the formation of parametrically driven DKSs. Further details of the experimental setup are provided in the Methods section.

As shown in Fig. 3a, we first investigated the comb dynamics by scanning the pump frequency from the blue to red detuning side at various pump powers. When the pump power was below the SHG saturation threshold of 25 mW, no cascaded quadratic nonlinear processes were observed, as indicated by the absence of sharp features in the cavity transmission of both the pump and SH. When the pump power exceeded the DQS threshold of 820 mW, step-like features appeared in the cavity transmission (Figure 3b), corresponding to distinct dynamical states in agreement with numerical simulations. As anticipated, stable, temporally confined bichromatic DQSs were generated at large negative pump phase detunings. Figure 3c presents the measured pump and SH optical spectra for representative dynamical states. Ultimately, at a pump phase detuning around $\delta_1 = -5\,\alpha_1$, a single DQS is generated with a 3-dB bandwidth of 1.15 THz at the pump and 1.13 THz at the SH (Figure 4).

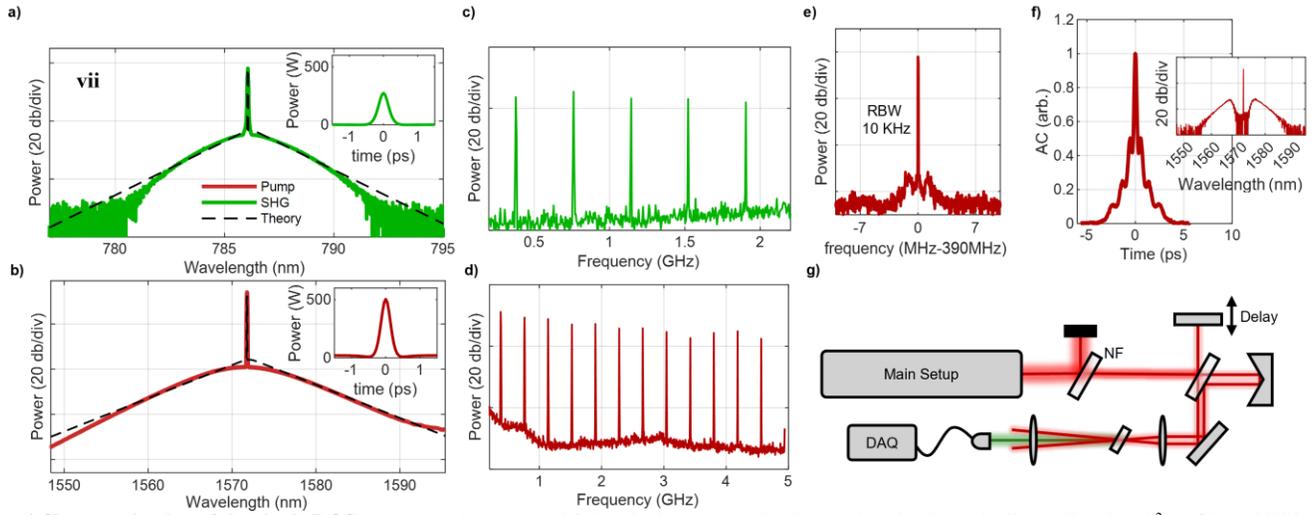

**Fig. 4 Characterization of the single DQS state** – a) The measured SH optical spectrum closely matches the theoretically predicted $sech^2$ profile, exhibiting a 3-dB bandwidth of 1.13 THz. The inset displays the corresponding temporal pulse shape with a 3-dB duration of 279 fs. b) The measured pump optical spectrum closely matches the theoretically predicted $sech^2$ profile, exhibiting a 3-dB bandwidth of 1.15 THz. The inset displays the corresponding temporal pulse shape with a 3-dB duration of 274 fs. c) The measured SH electrical spectrum is flat with clean beat notes spanning the photodetector-limited bandwidth of 2 GHz. d) The measured pump electrical spectrum is flat with clean beat notes spanning the photodetector-limited bandwidth of 5 GHz. e) The fundamental repetition rate beat note has a resolution-limited linewidth with a signal-to-noise ratio of 50 dB at a resolution bandwidth (RBW) of 10 KHz. f) The intensity autocorrelation (AC) measurement of the filtered pump DQS validates the generation of bright femtosecond pulses. Inset is the AC spectrum after the 12-nm notch filter. g) The schematic of the AC measurement setup. NF: notch filler. DAQ: data acquisition card.

**Characterization of the single DQS state.** Figures 4a and 4b display the measured bichromatic DQS optical spectra, which closely follow the theoretically predicted $sech^2$ profiles. The pump DQS exhibits a 3-dB bandwidth of 1.15 THz, corresponding to a transform-limited 3-dB pulse duration of 274 fs. The SH DQS features a 3-dB bandwidth of 1.13 THz, yielding a transform-limited 3-dB pulse duration of 279 fs. Flat electrical spectra with well-defined beat notes at the fundamental repetition rate and its harmonics were observed for both the pump and SH DQSs (Figures 4c and 4d), consistent with the generation of ultrashort pulses. Additionally, the coherence of the DQSs is verified by the narrow linewidth and high signal-to-noise ratio of the fundamental beat note, as shown in Figure 4e.

Finally, Fig. 4f shows the intensity autocorrelation (AC) measurement of the filtered pump DQS that validates the generation of bright femtosecond pulses. Although the 12-nm notch filter (NF) effectively suppressed the pump for the AC measurement (Figure 4g), its sharp spectral edges introduced pulse distortion, resulting in pronounced AC tails. Additional details are provided in Section VII of the Supplementary Information.

**Discussion**

In summary, we have demonstrated a DQS mode-locking of a continuous-wave pumped, doubly resonant cavity-enhanced second-harmonic generation system. By harnessing cascaded quadratic nonlinearities, we realize an effective Kerr nonlinearity that not only surpasses the intrinsic material Kerr nonlinearity by over three orders of magnitude but is also tunable in both magnitude and sign via pump phase detuning. This *in situ* engineered nonlinearity is unique in enabling femtosecond DQS generation in both free-space and chip-scale cavities, regardless of whether the dispersion is normal or anomalous. It offers a broad design space for soliton formation, enhancing flexibility in tailoring the soliton spectral range and supporting the possibility of universal soliton existence. The GVM between the pump and second harmonic fields acts as a primary perturbation to DQS dynamics, analogous to the role of third-order dispersion in DKS [28]. One strategy to mitigate GVM-induced distortion is to increase the total linear cavity loss for the pump, though this comes at the cost of a higher pump power threshold.

Numerical simulations predict distinct dynamical regimes depending on phase detuning, and experiments confirm the spontaneous emergence of bichromatic femtosecond DQSs spanning visible and near-infrared wavelengths. The observed DQSs exhibit spectral 3 dB bandwidths and transform-limited pulse durations of 1.15 THz and 274 fs at 1572 nm and 1.13 THz and 279 fs at 786 nm. These results introduce a robust and scalable route to nonlinear frequency conversion, mitigating the need for complex synchronization electronics, and ultrafast comb generation at unconventional wavelengths. This work deepens the understanding of dissipative soliton physics in quadratic systems and establishes a universal and practical platform for femtosecond pulse generation across diverse cavity platforms. The demonstrated flexibility and low operational threshold make DQSs promising for advancing quantum optics, frequency metrology, and integrated photonics.